# Towards more plausible point-identifying assumptions in two-sample Mendelian randomization

Fernando Pires Hartwig[1,2*], Neil Martin Davies[3,4,5], George Davey Smith[2,6]

[1]Postgraduate Program in Epidemiology, Federal University of Pelotas, Pelotas, Brazil.

[2]MRC Integrative Epidemiology Unit, University of Bristol, Bristol, UK.

[3]Division of Psychiatry, University College London, London, UK.

[4]Department of Statistical Sciences, University College London, London, UK.

[5]Department of Public Health and Nursing, Norwegian University of Science and Technology, Trondheim, Norway.

[6]Population Health Sciences, University of Bristol, Bristol, United Kingdom.

*Corresponding author. Postgraduate Program in Epidemiology, Federal University of Pelotas, Pelotas (Brazil) 96020-220. Phone: 55 53 32841300. E-mail: fernandophartwig@gmail.com.

**Abstract**

Two-sample Mendelian randomization (MR) is a widely used methodology in epidemiology. In two-sample MR, summary data (typically regression coefficients and standard errors) quantifying the association between multiple genetic variants and the exposure and the outcome are used in an instrumental variable framework to estimate the causal effect of the exposure on the outcome. Most two-sample MR methods assume that the association between each genetic variant and the exposure, as well as the causal effect of the exposure on the outcome, are constant on the additive scale. These assumptions are useful because they imply that, if all genetic variants were valid instruments, they would all estimate the same causal parameter – namely, the constant causal effect. We call this condition summary-level homogeneity. In this paper, we show that summary-level homogeneity follows from the following conditions: the causal effect is linear on the additive scale within population strata, but is not necessarily constant across strata; and heterogeneity in the causal effect is uncorrelated with heterogeneity in the association between each genetic variant and the exposure. Under these conditions, typical two-sample MR methods can be interpreted as estimators of the average causal effect. These results clarify that point-identifying assumptions required for two-sample MR methods are weaker than previously anticipated, making them more plausible and clarifying the interpretation of the resulting estimates if those assumptions hold.

## Introduction

Mendelian randomization (MR),[1] implemented within an instrumental variable (IV) framework in which genetic variants serve as IVs to estimate the causal effect of an exposure on an outcome, has become widely used in epidemiology.[2-4] This has been facilitated by methodological developments. In particular, two-sample MR can be implemented using only summary data (typically regression coefficients and standard errors) quantifying variant-exposure and variant-outcome associations.[5-7] Such data are widely available and incorporated into accessible tools.[8,9]

In many two-sample MR methods papers, it is assumed that, for each genetic variant, the variant-exposure association is constant on the additive scale across population strata (instrument strength homogeneity), although it may differ between variants. It is also often assumed that a one-unit increase in exposure produces the same additive change in outcome across strata (causal effect homogeneity). These assumptions concern variation within populations and are distinct from assumptions about comparability between the two source populations – i.e., the populations where variant-exposure and variant-outcome associations where estimated.

Even though homogeneity is a useful simplification, its plausibility is often questionable. For example, if the outcome is binary, a constant non-zero effect on the additive scale (a risk difference) is implausible as it would imply that the exposure fully determines the outcome. Moreover, if in addition the exposure is continuous, a constant non-zero risk difference assumption cannot hold over an unrestricted exposure range (although it may be a reasonable approximation over a limited range). Indeed, in such cases, a non-linear data-generating model such as a logistic model is typically assumed, in which case the exposure-outcome effect on the additive scale will generally depend on the starting level of the exposure, thus failing to be constant. Moreover, whether the exposure and outcome are continuous or binary, homogeneity of the variant-exposure association and of the exposure-outcome effect assumes there are no modifiers of the respective relationships on the additive scale, which cannot in general be taken for granted.[10,11]

In this paper, we show that weaker conditions are sufficient for two-sample MR estimators to identify a well-defined causal parameter.

## Summary-level homogeneity

Let $X$, $Y$ and $U$ respectively denote the exposure, outcome and common causes of $X$ and $Y$, assumed to be unmeasured. Let $G_1, \dots, G_L$ be a set of $L$ mutually independent genetic variants, indexed by $j \in \{1, \dots, L\}$. In the typical two-sample MR setting where $X$ and $Y$ are continuous, the population parameters quantifying the genetic associations are $\gamma_j^X = \mathrm{Cov}[X, G_j]/\mathrm{Var}[G_j]$ (the slope of the simple population linear regression of $X$ on $G_j$) and $\gamma_j^Y = \mathrm{Cov}[Y, G_j]/\mathrm{Var}[G_j]$ (the slope of the simple population linear regression of $Y$ on $G_j$).

Typically, two-sample MR methods assume the following model for $\gamma_j^Y$:

$$\gamma_j^Y = \gamma_j^X \varphi + \alpha_j \quad (1),$$

where $\varphi$ is the causal parameter of interest and $\alpha_j$ can be interpreted as the component of $\gamma_j^Y$ corresponding to violations of the assumptions required for $\varphi$ identification. Since the causal parameter that multiplies $\gamma_j^X$ is the same for all genetic variants, we say that summary-level homogeneity holds in (1). Summary-level homogeneity is a critical assumption of many two-sample MR methods, because it allows interpreting differences between instruments as evidence of $\alpha_j \neq 0$ for some $j$ – that is, that some genetic variants are invalid instruments. Heterogeneity in the effect estimates across different variants can be interpreted as violations of the instrumental variable assumptions (e.g., horizontal pleiotropy) only if this assumption holds – otherwise, heterogeneity could result from different instruments identifying different causal parameters.

**NOSH is sufficient for summary-level homogeneity**

Under instrument strength homogeneity (for each $j$) and causal effect homogeneity, summary-level homogeneity holds and $\varphi$ is the constant causal effect of $X$ on $Y$. To show that weaker assumptions are sufficient for summary-level homogeneity, initially assume every $G_j$ is a valid instrument. More specifically, assume the standard core IV assumptions:

Relevance: $\text{Cov}\left[G_j, X\right] \neq 0$ for all $j \in \{1, \ldots, L\}$ (2)

Independence: $G_j \perp U$ for all $j$ (3)

Exclusion restriction: $G_j \perp Y \big| (U, X)$ for all $j$ (4)

Now, make the following additional assumptions:

$$\mathrm{E}[Y^{X=x} | U = u] = \beta_0(u) + \beta_Y(u)x \quad (5)$$

$$\text{Cov}\left[\gamma_j^X(U), \beta_Y(U)\right] = 0 \quad (6),$$

where $Y^{X=x}$ is the potential outcome under exposure $X = x$; and $\gamma_j^X(U) = \text{Cov}\left[X, G_j | U\right] / \text{Var}\left[G_j \big| U\right]$. In words, (5) states that the average effect of $X$ on $Y$ is, within strata of $U$, linear on the additive scale. We also assume consistency, so $X = x \Rightarrow Y = Y^{X=x}$. These assumptions collectively amount to the NO Simultaneous Heterogeneity (NOSH) assumption.[12]

If an IV is valid and NOSH holds, then the Wald estimand equals the average causal effect (ACE)[12] (proof available in the supplement). Given the notation above, the ACE is $\mathrm{E}[\beta_Y(u)]$. Therefore, $\theta_j = \mathrm{E}[\beta_Y(U)]$ for all $j$. Since $\theta_j = \gamma_j^Y / \gamma_j^X$, this implies $\gamma_j^Y = \gamma_j^X \mathrm{E}[\beta_Y(U)]$.

Therefore, assumptions (2-6) are sufficient for $\gamma_j^Y = \gamma_j^X \mathrm{E}[\beta_Y(U)]$. This implies that, if $\gamma_j^Y \neq \gamma_j^X \mathrm{E}[\beta_Y(U)]$, then at least one of the assumptions (3-6) is violated (assumption (2) is assumed to hold since it is empirically verifiable). We can therefore allow for some genetic variants to be invalid IVs by letting:

$$\gamma_j^Y = \gamma_j^X \mathrm{E}[\beta_Y(U)] + \alpha_j \quad (7),$$

where $\alpha_j$ can be interpreted as the component of $\gamma_j^Y$ corresponding to violations of at least one of assumptions (3-6). Note that (7) satisfies summary-level homogeneity because it has the same structure as (1) with $\varphi = \mathrm{E}[\beta_Y(U)]$. Therefore, the above assumptions are sufficient for interpreting two-sample MR methods assuming (1) as estimators of the ACE, provided their specific identification conditions hold.

**Interpreting common two-sample MR methods under NOSH**

The model in (7) implies $\theta_j = \mathrm{E}[\beta_Y(U)] + b_j$, where $b_j = \alpha_j / \gamma_j^X$ is the bias parameter of $\theta_j$. This expression for $\theta_j$, involving a common causal parameter and an instrument-specific bias parameter, is the typical setup of MR estimators that combine different ratio estimates in different ways.

For example, the estimand of the inverse-variance weighted (IVW) method[5] is $\pi_{\text{IVW}} = \sum_{j=1}^{L} \theta_j \omega_j = \mathrm{E}[\beta_Y(U)] + \sum_{j=1}^{L} b_j \omega_j$, where $\omega_j > 0$ satisfying $\sum_{j=1}^{L} \omega_j = 1$ is the weight $\theta_j$ receives. Clearly, $\pi_{\text{IVW}} = \mathrm{E}[\beta_Y(U)]$ if and only if $\sum_{j=1}^{L} b_j \omega_j = 0$. A sufficient condition for this is to assume that all genetic variants are valid IVs, so that $\alpha_j = b_j = 0 \; \forall \, j$. Similarly, the identification conditions of the median and the mode would remain the same: $\text{median}(b_j) = \text{mode}(b_j) = 0$ for the unweighted estimators, and analogous weighted versions of the identification conditions for the weighted estimators.[13,14] Under these identification conditions, $\text{median}(\theta_j) = \text{median}(\text{E}[\beta_Y(U)] + b_j) = \text{E}[\beta_Y(U)]$ and $\text{mode}(\theta_j) = \text{mode}(\text{E}[\beta_Y(U)] + b_j) = \text{E}[\beta_Y(U)]$.

Other two-sample MR methods do not rely on combining ratio estimates. The most common one is MR-Egger regression,[15] in which $\gamma_j^Y$ is regressed on $\gamma_j^X$ (typically, under some weighting scheme, ignored here for simplicity). In this case, the MR-Egger estimand is $\pi_{\text{Egger}} = \text{Cov}[\gamma_j^X, \gamma_j^Y]/\text{Var}[\gamma_j^X]$. Since $\text{Cov}[\gamma_j^X, \gamma_j^Y] = \text{Cov}[\gamma_j^X, \gamma_j^X \text{E}[\beta_Y(U)] + \alpha_j] = \text{E}[\beta_Y(U)]\text{Var}[\gamma_j^X] + \text{Cov}[\gamma_j^X, \alpha_j]$, we have that $\pi_{\text{Egger}} = \text{E}[\beta_Y(U)] + \text{Cov}[\gamma_j^X, \alpha_j]/\text{Var}[\gamma_j^X]$. Under the assumption that $\text{Cov}[\gamma_j^X, \alpha_j] = 0$ (often referred to as the Instrument Strength Independent of Direct Effect – InSIDE – assumption), we have that $\pi_{\text{Egger}} = \text{E}[\beta_Y(U)]$.

## Discussion

We showed that NOSH implies the summary-level homogeneity condition typically assumed by two-sample MR methods. More specifically, NOSH allows interpreting these methods as estimators of the ACE, a well-defined causal parameter that is often of substantive interest. Assuming NOSH holds for all genetic variants seems more promising for two-sample MR (at least with respect to currently used methods) than assuming monotonicity – i.e., $\gamma_j^Y \geq 0$. This is because monotonicity does not ensure that the estimand is the same across IVs, and therefore insufficient for summary-level homogeneity.

In the formulation above, NOSH violations can contribute to $\alpha_j$. This is not typical in the MR literature, where terms representing “direct effect” (in a broad sense) arises only fromviolations of the core IV assumptions. However, this is only because homogeneity conditions are assumed to hold, rather than because of an inherent limitation of NOSH relative to homogeneity. Indeed, if NOSH holds, then in the above formulation the $\alpha_j$ parameter would also correspond only to violations of the core IV assumptions. Moreover, it is natural that NOSH violations contribute to $\alpha_j$, since such violations (like violations of the other IV assumptions) would generally bias two-sample MR methods as estimators of the ACE.

Our results are more applicable to two-sample MR when both $X$ and $Y$ are continuous, since in this case summary genetic associations are typically quantified as linear regression coefficients. Indeed, NOSH is more plausible for continuous $X$ and $Y$ because non-linear data-generating models (e.g., logistic) are typically assumed for binary variables, in which case condition (6) (and condition (5) if $Y$ is binary and $X$ is continuous) is less plausible. Further methodological research is needed on point-identifying assumptions sufficient for two-sample MR methods to be well-defined when $X$ and/or $Y$ are binary. Even if both $X$ and $Y$ are continuous, linearity of causal effect of $X$ on $Y$ within population strata may not be plausible in some practical applications. Without linearity, our results do not hold and two-sample MR estimates are ill-defined.[16]

Although NOSH is weaker than homogeneity, it can still be false. Its plausibility can be assessed by empirical data on the existence and magnitude of effect modification on the additive scale. This can be done on a large scale, for example, through genome-wide scans on multiple continuous traits that assess evidence for genetic effect heterogeneity via variance tests,[17-19] as well as on a case-by-case basis involving theory-driven assessment of candidate effect modifiers.[11,20] NOSH also requires the consistency assumption. Assessing the plausibility of this assumption (in addition to the IV assumptions that are routinely discussed in MR papers) can be aided by careful consideration of the gene-environment equivalence principle.[21,22]. Our results should improve the interpretation of two-sample MR estimates when the NOSH assumption is plausible.

**Source of funding:** The Medical Research Council (MRC) and the University of Bristol support the MRC Integrative Epidemiology Unit (MC_UU_00032/01). FPH is supported by a Research Productivity Fellowship from the Brazilian National Council for Scientific and Technological Development (grant no. 303880/2023-6). NMD is supported via a Norwegian Research Council Grant (295989) and the UCL Division of Psychiatry (https://www.ucl.ac.uk/psychiatry/division-psychiatry).

**Conflicts of interest:** None declared.

**Dedication:** This work is dedicated to the memory of Bento/Clara (FPH's unborn child).

**Supplementary material**

Here we prove that assumptions (2)-(6) are sufficient for the Wald estimand to equal the average causal effect.

$$\mathrm{E}\left[Y\middle|G_j,U,X=x\right]=\mathrm{E}[Y|U,X=x] \quad \text{(by (4))}$$
$$=\mathrm{E}[Y^{X=x}|U,X=x] \quad \text{(by consistency)}$$
$$=\mathrm{E}[Y^{X=x}|U] \quad \text{(by exchangeability)}$$
$$=\beta_0(U)+\beta_Y(U)x \quad \text{(by (5))} \quad (\mathrm{S1})$$

$$\mathrm{E}\left[Y\middle|G_j,U\right]=\mathrm{E}\left\{\mathrm{E}\left[Y\middle|X,G_j,U\right]\middle|G_j,U\right\}$$
$$=\mathrm{E}\left\{\beta_0(U)+\beta_Y(U)X\middle|G_j,U\right\} \quad \text{(by (S1))}$$
$$=\beta_0(U)+\beta_Y(U)\mathrm{E}\left[X\middle|G_j,U\right] \quad (\beta_0(U)\text{ and }\beta_Y(U)\text{ constant conditional on }U) \quad (\mathrm{S2})$$

$$\mathrm{Cov}\left[Y,G_j|U\right]=\mathrm{Cov}\left\{\mathrm{E}\left[Y\middle|G_j,U\right],G_j|U\right\}$$
$$=\mathrm{Cov}\left\{\beta_0(U)+\beta_Y(U)\mathrm{E}\left[X\middle|G_j,U\right],G_j|U\right\} \quad \text{(by (S2))}$$
$$=\beta_Y(U)\mathrm{Cov}\left\{\mathrm{E}\left[X\middle|G_j,U\right],G_j|U\right\} \quad (\beta_0(U)\text{ and }\beta_Y(U)\text{ constant conditional on }U)$$
$$=\beta_Y(U)\mathrm{Cov}\left[X,G_j|U\right]$$
$$=\beta_Y(U)\gamma_j^X(U)\mathrm{Var}\left[G_j\middle|U\right] \quad (\gamma_j^X(U)=\mathrm{Cov}\left[X,G_j|U\right]/\mathrm{Var}\left[G_j\middle|U\right])$$
$$=\beta_Y(U)\gamma_j^X(U)\mathrm{Var}\left[G_j\right] \quad \text{(by (3))} \quad (\mathrm{S3})$$

$$\mathrm{Cov}\left[Y,G_j\right]=\mathrm{E}\left[\mathrm{Cov}\left[Y,G_j|U\right]\right]$$
$$=\mathrm{E}\left[\beta_Y(U)\gamma_j^X(U)\mathrm{Var}\left[G_j\right]\right] \quad \text{(by (S3))}$$
$$=\mathrm{E}\left[\beta_Y(U)\gamma_j^X(U)\right]\mathrm{Var}\left[G_j\right]$$
$$=\mathrm{E}[\beta_Y(U)]\mathrm{E}\left[\gamma_j^X(U)\right]\mathrm{Var}\left[G_j\right] \quad \text{(by (6))} \quad (\mathrm{S4})$$

$$\mathrm{Cov}\left[X,G_j\right]=\mathrm{E}\left\{\mathrm{Cov}\left[X,G_j|U\right]\right\}+\mathrm{Cov}\left\{\mathrm{E}[X|U],\mathrm{E}\left[G_j|U\right]\right\}$$
$$=\mathrm{E}\left\{\mathrm{Cov}\left[X,G_j|U\right]\right\}+\mathrm{Cov}\left\{\mathrm{E}[X|U],\mathrm{E}\left[G_j\right]\right\} \quad \text{(by (3))}$$
$$=\mathrm{E}\left\{\mathrm{Cov}\left[X,G_j|U\right]\right\}$$

$$= \mathrm{E}\left\{\gamma_j^X(U)\mathrm{Var}\left[G_j\middle|U\right]\right\}$$ ($\gamma_j^X(U) = \mathrm{Cov}\left[X, G_j|U\right]/\mathrm{Var}\left[G_j\middle|U\right]$)

$$= \mathrm{E}\left[\gamma_j^X(U)\right]\mathrm{Var}\left[G_j\right]$$ (by (3)) (S5)

$$\theta_j = \mathrm{Cov}\left[Y, G_j\right]/\mathrm{Cov}\left[X, G_j\right]$$

$$= \left(\mathrm{E}[\beta_Y(U)]\mathrm{E}\left[\gamma_j^X(U)\right]\mathrm{Var}\left[G_j\right]\right)/\left(\mathrm{E}\left[\gamma_j^X(U)\right]\mathrm{Var}\left[G_j\right]\right)$$ (by (S4) and (S5))

$$= \mathrm{E}[\beta_Y(U)]. \square$$